\documentclass[12pt,a4paper,final]{iopart}    

\usepackage[dvips]{graphicx}  
\usepackage[latin1]{inputenc} 
\usepackage{amssymb}          
\usepackage{amsfonts}
\usepackage{latexsym}
\usepackage[usenames,dvipsnames]{xcolor}
\usepackage{curves}
\usepackage{comment}

\newtheorem{thm}{Theorem}[section] 		
\newtheorem{prop}{Proposition}[section] 	

\newtheorem{defin}{Definition}

\newcommand{\proof}{\noindent {\em Proof. }}

\newcommand{\ket}[1]{|#1\rangle}
\newcommand{\bra}[1]{\langle #1|}

\newcommand{\Hi}{\mathcal{H}}

\newcommand{\trace}{\mathrm{{Tr}}}

\newcommand{\Ec}{\mathcal{E}}
\newcommand{\Tc}{\mathcal{T}}

\newcommand{\Ti}{\mathcal{T}}

\renewcommand{\span}{\textrm{span}}
\newcommand{\rank}{\textrm{rank}}

\newcommand{\qed}{\hfill $\Box$ \vskip 2ex}

\newcommand{\eqref}[1]{(\ref{#1})}

 \newcommand{\beq}{\begin{equation}}
 \newcommand{\eeq}{\end{equation}}
 \newcommand{\beqa}{\begin{eqnarray}}
 \newcommand{\eeqa}{\end{eqnarray}}
 \newcommand{\beqan}{\begin{eqnarray*}}
 \newcommand{\eeqan}{\end{eqnarray*}}
 \newcommand{\bea}{\begin{eqnarray}}
 \newcommand{\eea}{\end{eqnarray}}
 
\date{\today}

\begin{document}

\title[Unitary design of quantum channels]
{Quantum and classical resources for unitary design of open-system evolutions}

\author{Francesco Ticozzi$^{1,2}$, Lorenza Viola$^2$}          
\address{$^1$ Dipartimento di Ingegneria dell'Informazione,
Universit\`a di Padova, \\ via Gradenigo 6/B, 35131 Padova, Italy}
\address{$^2$ Department of Physics and Astronomy, Dartmouth 
College, \\ 6127 Wilder Laboratory, Hanover, NH 03755, USA}
\ead{ticozzi@dei.unipd.it, lorenza.viola@dartmouth.edu}

\begin{abstract} 
A variety of tasks in quantum control, ranging from purification and cooling, to quantum stabilization and 
open-system simulation, rely on the ability to implement a target quantum channel 
over a specified time interval within prescribed accuracy. This can be achieved by engineering a suitable unitary dynamics of the 
system of interest along with its environment -- which, depending on the available level of control, is fully or partly exploited as a 
{\em coherent quantum controller}. 
After formalizing a controllability framework for completely positive trace-preserving quantum dynamics,  
we provide sufficient conditions on the {\em environment state and dimension} that allow for the realization of relevant classes of quantum 
channels -- including extreme channels, stochastic unitaries, or simply any channel. The results hinge on generalizations of Stinespring's 
dilation via a {\em subsystem principle}. In the process, we show that a conjecture by Lloyd on the minimal dimension of the environment 
required for arbitrary open-system simulation, albeit formally disproved, can in fact be salvaged -- provided that 
{\em classical randomization} is included among the available resources. 
Existing measurement-based feedback protocols for universal simulation, dynamical decoupling, and dissipative state preparation are 
recast within the proposed coherent framework as concrete applications, and the resources they employ discussed in the light of 
the general results.
\end{abstract}




\section{Introduction}
Realistic physical systems are never perfectly isolated from their surrounding environment -- due to both unwanted couplings to 
uncontrolled degrees of freedom and to designer interactions with measurement apparatuses or auxiliary controller devices. 
In the statistical description of quantum systems, the resulting class of {\em open-system dynamics}
may be derived directly from the quantum mechanics postulates \cite{kraus}.  
More precisely, the state of the target system is associated to a trace-one, positive semidefinite density operator and, 
under the assumption that no initial correlations are present with the environment, 
its evolution over some specified time interval is described by a {\em completely positive, trace-preserving} (CPTP) linear map. 
Physically, the latter results from averaging over the degrees of freedom of the environment after a unitary evolution, driven by a joint 
Hamiltonian, has taken place.
Beside their natural emergence in quantum statistical mechanics, open-system theory and thermodynamics \cite{davies,petruccione,streater}, 
CPTP dynamics have gained a central role within quantum information science \cite{nielsen-chuang}. On the one hand, CPTP maps 
are the natural non-commutative analogues of classical stochastic maps; as such, they are being widely used to model 
quantum communication channels, noise effects, quantum error-correcting procedures, erasure and reset operations. 
In this context, they are typically called {\em quantum channels}, and we will use here the two denominations interchangeably. On the 
other hand, CPTP evolutions play a pervasive role also in quantum measurement theory and statistics -- in particular, describing 
generalized non-selective measurements, conditional expectations and quantum filters 
\cite{petz-book,Barc09,belavkin-filtering}, as well as feedback networks in quantum control theory 
\cite{WM,altafini-tutorial,wiseman-milburn}.

In this work, we focus on the issue of \emph{quantifying the resources needed to engineer}, exactly or within a finite accuracy, 
a desired CPTP map. 
Our interest in this problem stems from two major motivations. On a fundamental level, it is a key theoretical issue in the design of 
{\em universal, digital open-system simulators} \cite{lloyd-science,viola-engineering} --
one of the premier applications of quantum information science, and one in which rapid experimental progress is being made  
\cite{universaldigitalsimulation,blatt-maps}. 
In addition, a variety of key tasks in quantum control can be described as, or can be brought to bear on, 
the effective engineering of a target set of CPTP maps: among these, we mention dynamical decoupling, 
quantum stabilization, purification and cooling -- as we also investigated in previous work 
\cite{viola-dd,ticozzi-markovian,rapidpurification,ticozzi-cooling}. 

It has long been known, thanks to a representation theorem by Stinespring \cite{Stinespring1955}, that {\em any} CPTP map can 
in principle be obtained via a {\em unitary dilation}. More concretely, this entails pairing the target system $S$, say, of dimension 
$d_S,$ to an auxiliary system $E$, with dimension {\em at most} $d_S^2$ and prepared in a known {\em pure} state, 
and then implementing a joint unitary evolution on $S + E$, whose net effect on $S$ is to enact the target map. 
However, while this provides a sufficient set of resources, characterizing what resources may also be necessary is 
not straightforward. In particular, it is not a priori clear what minimal dimension of $E$ is needed to implement any target 
map through such a \emph{unitary design}, nor the extent to which access to a {\em mixed} initial state of $E$ may 
hinder the task. 
If one relaxes the problem to one of approximate engineering of a target map within a prescribed tolerance, 
even sufficient conditions are lacking to the best of our knowledge. 

In addressing these issues, we introduce a general system-theoretic scenario for coherent control of open systems, and define a set of 
relevant {\em CPTP controllability notions}, associated, respectively, to the ability of engineering: (i) the full convex set of CPTP maps; or 
(ii) only the extreme ones; or (iii) all of those of fixed Kraus rank -- either exactly or within prescribed non-zero accuracy (Section \ref{sec:preliminaries}). We then proceed to derive a series of sufficient and/or necessary conditions for these controllability notions to 
hold (Sections \ref{sec:controllability} \& \ref{sec:maj}).  
These results are obtained by generalizing Stinespring dilations, as well as previous results specifically regarding purification and cooling 
\cite{brumer-cooling,ticozzi-cooling}: in particular, we provide a {\em sufficient condition} on (possibly mixed) environment states that ensure unitary 
engineering of any maps of limited  Kraus rank within a prescribed accuracy $\varepsilon >0$. This condition amounts to the existence of 
a sufficiently, $\varepsilon$-pure state in a {\em virtual-subsystem decomposition} of the environment, stemming from the ``subsystem principle''  established for purifying quantum maps in Ref. \cite{ticozzi-cooling}. Next, we show that the same conditions are also {\em necessary} 
for unitary design of the set of extreme maps. However, the existence of an approximately pure quantum subsystem turns out {\em not} to be 
necessary in general for engineering specific target CPTP maps -- as we explicitly demonstrate for {\em stochastic unitaries} via a 
construction based on majorization (Section \ref{sec:stochuni}). While the controllability results we provide are not constructive, 
at least in terms of making reference to {\em specific} control resources at hand, we propose a way to recast the unitary 
engineering of a CPTP map as an optimal {\em state steering problem} 
by a direct application of the channel-state Choi-Jamio{\l}kowski duality \cite{choi}. 

It is worth recalling that in his work on universal quantum simulators \cite{lloyd-science}, Lloyd conjectured that exact open-system simulation 
could be realized even with the dimension of the auxiliary system being reduced from $d_S^2$ to $d_S,$ provided that the latter could be 
prepared in an arbitrary (pure or mixed) quantum state. A number of explicit counterexamples have subsequently disproved the validity of this conjecture \cite{terhal,other,other1}. Interestingly, we find that a version of the above conjecture \emph{does} hold true, provided that additional 
{\em classical} randomization resources, as well as non-deterministic channel constructions, are allowed (Section \ref{sec:prob}). 
Specifically, we show that any target, non-extreme map may be obtained as the {\em average} over a randomized set of 
extreme-map dilations, so that any CPTP is reachable by using an auxiliary system that is, indeed, just $d_S$-dimensional. 
With respect to Lloyd's original conjecture, we need only pure state of the auxiliary system, but we allow for sampling from an 
arbitrary {\em classical} distribution on a larger space -- one whose cardinality may be up to $d_S^4$. 

Our study bears similarities, as well as fundamental differences, with the analysis of {\em indirect controllability}, in the language 
of \cite{dalessandro-indirect}. A first difference is that the task is therein limited to the engineering of {\em unitary} evolutions on the 
target system. In addition, our necessary conditions for the engineering of extreme maps show that it is impossible, for general non-unitary 
evolutions as we consider, to have CPTP controllability independently of the state of $E$. Our work also complements existing results on 
controllability of open-system Markovian dynamics, including continuous-time semigroups \cite{altafini-open,thomas-controllability} and 
discrete-time dynamics \cite{kraus-controllability,albertini-feedback}. 

Thanks to the flexible framework we employ, our results may be applied and specialized to a number or existing protocols for 
universal simulation of CPTP dynamics or for synthesizing specific CPTP maps of interest.
In the last part of the paper (Section \ref{sec:examples}), we specifically re-examine three such applications 
within our framework -- namely, using only coherent Hamiltonian evolutions and coherent quantum feedback.
The first application is a constructive approach for simulating arbitrary CPTP maps to arbitrary accuracy, based on 
binary (``Yes-No'') measurements, proposed in Ref. \cite{viola-engineering}. 
While its original formulation employs only a single auxiliary qubit, the control resources 
also include the ability of resetting it to a known pure state. Here, we provide a fully coherent implementation of the protocol, 
by examining what resources are needed in this case, as well as the impact of having a mixed ancillary state. 
As a second illustrative application, we recast in fully coherent picture the {\em feedback decoupling} protocol we 
proposed in Ref. \cite{ticozzi-feedbackDD}: in this case, the task is to engineer a trivial evolution (an effective ``NOOP gate''  
on the target system) by averaging out the effect of an uncontrollable bath.
Lastly, we extend the {\em splitting-subspace} approach for stabilizing a quantum state in finite time introduced 
in Ref. \cite{baggio-CDC}, by allowing for a larger auxiliary space and, again, fully coherent resources --  
which guarantees the desired state or subspace stabilization to be achievable by a single iteration of the protocol.


\section{Preliminaries}\label{sec:preliminaries}

\subsection{Hamiltonian description of controlled open quantum systems}
\label{sec:prel}

\begin{figure}[t]
\centering
\hspace*{10mm}\includegraphics[width=0.5\columnwidth]{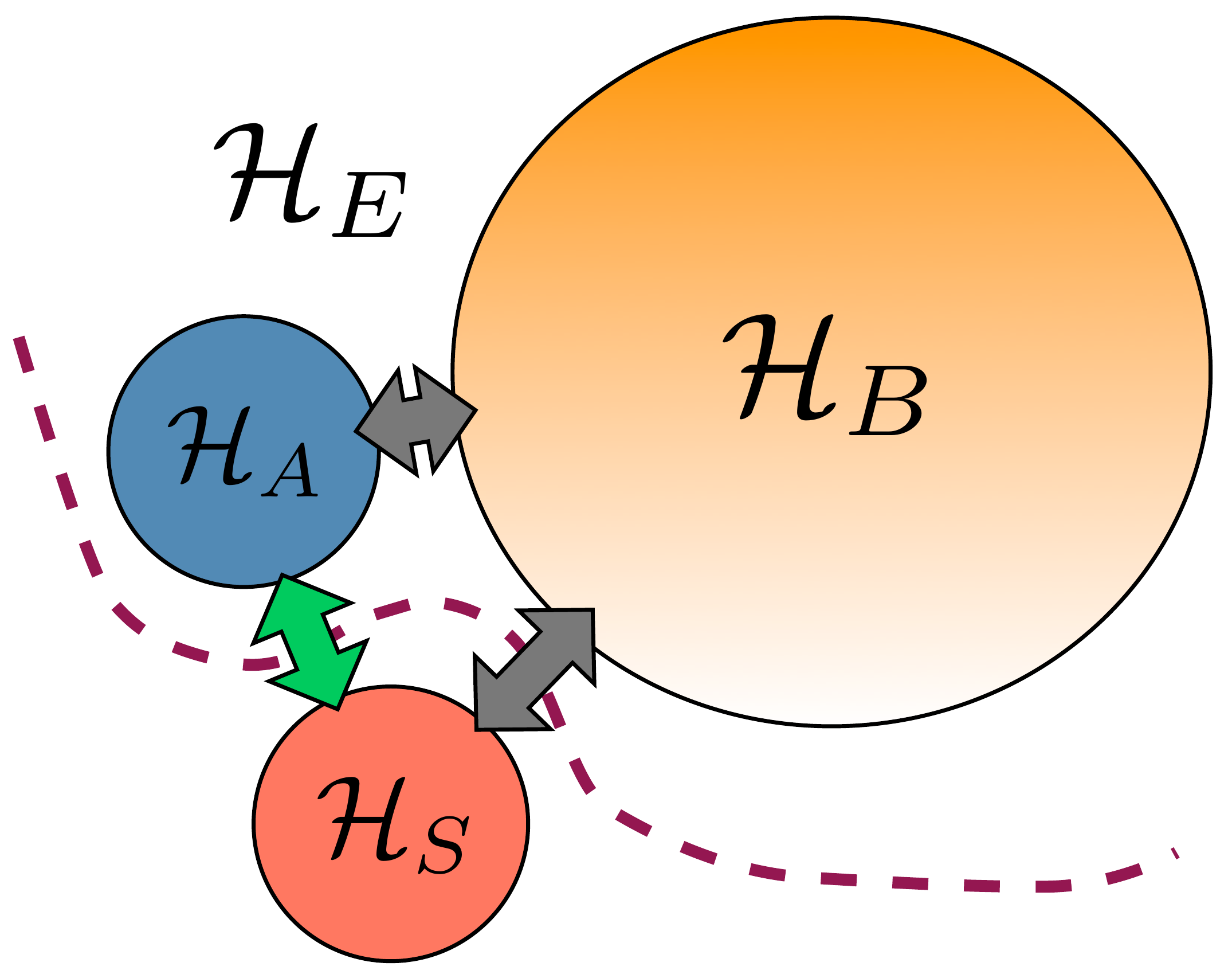}
\vspace*{-1mm}
\caption{(Color online) Tripartite setting of interest: $\Hi_S$ is the $d_S$-dimensional target system, 
$\Hi_B$ the uncontrollable bath, and $\Hi_A$ an engineered auxiliary system. {We assume 
both $S$ and $A$ to be finite-dimensional, whereas $E$ may be infinite-dimensional.}} 
\label{fig:tripartite}
\end{figure}

We consider an open-system framework that is flexible enough to include arbitrary control 
protocols for quantum dynamical engineering and simulation 
using semi-classical open-loop control and coherent feedback capabilities.
We assume that the system of interest, $S$, may be coupled to both an uncontrollable quantum
bath, $B$, via a {\em fixed} interaction Hamiltonian, as well as to an engineered auxiliary quantum system, 
$A$, via a tunable interaction. The latter may also couple to $B$ in general.
We shall refer to the pair $A,B$ collectively as the {\em environment}, $E$ (see also Fig. \ref{fig:tripartite}). 

Let $\Hi_S, \Hi_A, \Hi_B$ denote the Hilbert space of system, ancilla, and physical bath, respectively, 
with ${\rm{dim}}(\Hi_S)\equiv d_S, {\rm{dim}}(\Hi_A) \equiv d_A$, and ${\rm{dim}}(\Hi_E) \equiv d_E$ being 
defined accordingly. The initial state on the joint state space $\Hi_{SE}\equiv \Hi_S \otimes \Hi_E = \Hi_S \otimes
(\Hi_A \otimes \Hi_B)$ at a reference time $t=0$ is assumed to be  {\em factorized} with respect
to the tri-partition
$\rho_{SE}=\rho_S\otimes\rho_E=\rho_S\otimes \rho_{A}\otimes \rho_{B}$.
We take the controlled joint dynamics to be generated by a Hamiltonian of the form 
\beq 
\label{eq:syscon}
H(t) \equiv H_0+H_c(t) = H_S \otimes {I}_E + {I}_S \otimes
H_E + H_{SE}  + H_c(t),
\eeq 
where $H_S$ and $H_E \equiv H_A\otimes I_B + I_A \otimes H_B + H_{AB}$ 
account for the free Hamiltonians of $S$ and $E$ alone, 
$H_{SE} \equiv H_{SA}+H_{SB}$ includes the fixed interaction terms,
and  the control Hamiltonian acts trivially on the uncontrollable component $B$, that is, 
\beq
\label{eq:Hc}
H_c(t) \equiv \sum_\ell u_\ell(t) H_{\ell} \otimes I_B ,
\eeq
with the (real) functions $\{ u_\ell (t)\}$ being the control inputs and the control Hamiltonians
having the general form $H_\ell \equiv H_{S, \ell} \otimes I_A + I_S \otimes H_{A,\ell} + H_{SA, \ell}$. 
Note that the case where $B$ represents a classical bath may be formally included by moving to 
the interaction picture with respect to $H_B$ and replacing time-dependent bath operators with 
classical random variables.  

After a time $T> 0,$ the dynamics generated by $H(t)$ is described 
by the conjugate action of a unitary operator $U_{SE}(T)$,
belonging to the unitary group $\mathfrak{U}(\Hi_{SE}).$  
Since we start from factorized initial conditions and a fixed environment state, 
the {\em reduced state} of the system at time $T$ is a linear function of the initial system state, 
that is, we may write $\rho_S(T) \equiv \Ec_{T,0}(\rho_S)$, with
\beq
\Ec_{T,0}(\rho_S)  = 
\trace_E [U_{SE}(T)\,\rho_S\otimes\rho_E \, U_{SE}^\dag(T)]{ .}
\label{eq:reduced}
\eeq 
\noindent 
Thus, a target CPTP map $\Ti$ may be obtained via {\em unitary design at time $T$, from environment state $\rho_E$}, 
if, by using suitable controls, we can enact a joint unitary $U_{SE}(T)$ such that $\Ec_{T,0}=\Ti.$

 A well-known result by Kraus \cite{kraus} states that a linear map ${\cal E}$ is CPTP if and only if it admits an 
{\em operator-sum representation} (OSR), namely:  
\beq
\label{eq:kraus}
{\cal E} (\rho) = \sum_{j=1}^m M_j\rho M^\dag_j,
\quad \sum_{j=1}^m M_j^\dag M_j = I, \quad \forall \rho\in {\cal D}(\Hi_S), 
\eeq
in terms of the so-called Kraus operators $\{M_j\}$, and with ${\cal D}(\Hi_S)$ denoting the set of 
density operators on $\Hi_S$. The above discrete-time dynamics may be seen as the Hilbert-Schmidt dual of an 
(Heisenberg-picture) map $\Ec^\dag$ which is CP and unital, namely, a CP map that preserves the identity.
It is straightforward to show  that the map ${\cal E}_{T,0}$ in Eq. \eqref{eq:reduced} can be indeed represented as 
an operator sum, and hence it is CPTP \cite{nielsen-chuang}.

While the OSR is not unique, the minimal number of Kraus operators $m,$ which is 
called the {\em Kraus rank} of $\Ec$, is well defined, and always $m<d_S^2$ \cite{kraus,nielsen-chuang}. 
In order to define the Kraus rank 
precisely, it is convenient to introduce a different representation of CPTP maps, the so-called {\em 
Choi-Jamio{\l}kowski isomorphism} between quantum channels and states. Let $\Hi_{S'}$ be an isomorphic copy of the system's state space $\Hi_{S}.$ Choose a reference basis $\{\ket{\phi_i}\},\{\ket{\phi'_i}\}$ in each and 
consider the maximally entangled state $\Phi \equiv (\sum_i \ket{\phi_i}\ket{\phi'_i})(\sum_i\bra{\phi_i}\bra{\phi'_i})/d_S.$
Then, the {\em Choi matrix} \cite{choi} is defined by
\beq
\label{eq:choi} 
C_{\Ec}\equiv (\Ec\otimes {\cal I}_{S'}) (\Phi), 
\eeq
where ${\cal I}$ is the identity map.
It is possible to show that if two CPTP maps have the same Choi matrix, they are the same map, and the 
Kraus rank of $\Ec$ can be uniquely defined as  the rank of its associated Choi matrix \footnote{If $\Hi_S$ is countably infinite, 
then one can partially by-pass the difficulty arising from the normalization $1/d_S$ by using an unnormalized 
version of $\Phi.$}.

From the Kraus representation in Eq. (\ref{eq:kraus}), it is easy to see that CPTP maps form a convex set: 
if $\Ec_1$ and $\Ec_2$ are CPTP, then clearly $\lambda \Ec_1 + (1-\lambda) \Ec_2$ is CPTP for 
$\lambda\in[0,1].$ In the work of Choi \cite{choi}, a useful characterization of {\em extreme points} of the 
convex CPTP set is provided as well: a map is extreme if and only if (any of) its Kraus representation $\{M_k\}$ is such 
that the set of operators:
\beq
\label{eq:extreme} \{M_k^\dag M_j\}_{j,k=1}^m 
\eeq
are linearly independent. This is property is invariant with respect to the allowed changes of 
representation in the $\{M_k\}$. It follows immediately, by comparing dimensions, that extreme points 
have at most Kraus rank $m=d_S$.

{\em Remark:} We choose to work in terms of a controlled Hamiltonian setting for the dynamics, as 
opposed to a reduced dynamics description which is typical, for instance, in master equation approaches to
continuous-time dynamics, for a number of reasons.  First, 
this allows us to pinpoint the role of the environmental degrees of freedom on the attainable set of dynamics, 
while providing access to effectively non-Markovian evolution -- which is harder to describe in full generality 
at the level of reduced dynamics. Most importantly, the Hamiltonian setting does not restrict us to 
engineering of {\em divisible} CPTP maps only, which are known to be a strict subset of all possible ones \cite{wolf-dividing}. 
This is at variance with the capabilities of control protocols that entail sequences of ``elementary'' 
CPTP building blocks, arising either from exponentials of (generally) time-dependent Lindblad generators 
(hence yielding infinitesimal divisible channels) or from discrete-time dissipative quantum circuits 
(also accessing divisible but not necessarily infinitesimal divisible quantum channels) -- see e.g. 
\cite{Bacon2001,thomas-controllability,Petruccione2015,Sanders2013,Sanders2015,CamposV2016,fts-paper} 
for illustrative contributions. 

\subsection{Reachability definitions and control assumptions}

In the following, we will refer to the available {\em control resources} $(\{H_c(t) \},\rho_A)$, together with the fixed Hamiltonians 
$H_S,H_E,H_{SE}$ and bath state $\rho_B$, as a {\em control scenario}. Part of our results can be framed as controllability 
results, where one looks at the set of maps that can be enacted, or {\em reached}, within a specified control scenario. Formally:

\noindent
\begin{defin}{\bf (CPTP reachability by unitary design)}
A CPTP map $\Tc$ on $S$ is {\em reachable at time $T$} 
by unitary design in the control scenario $(\{H_c(t)\},\rho_A ; \rho_B)$ if 
we can enact a unitary $U_{SE}(T)$ so that in Eq. \eqref{eq:reduced} we have $\Ec_{T,0}=\Tc$. 
Likewise, $\Tc$ is {\em $\varepsilon$-approximate reachable at time $T$} if we can enact a unitary 
$U_{SE}(T)$ so that 
\beq 
\label{eq:eps}
d_{\rm TV}(\Ec_{T,0}(\rho),\Tc(\rho))\leq \varepsilon, \quad \forall \rho\in\mathfrak{D}(\Hi_S).
\eeq 
\end{defin}

\noindent 
Here, $d_{\rm TV}(X,Y)\equiv \frac{1}{2} \trace(|X-Y|) = \frac{1}{2}
\vert\vert X-Y \vert\vert_1$ is the quantum total-variation distance, which is a natural 
measure of distinguishability between quantum states 
\cite{nielsen-chuang,viola-IPS,ticozzi-isometries}. Equation \ref{eq:eps} is equivalent to 
requesting that the distance between the maps is small in the induced operator norm:
\[ \frac{1}{2}\|\Ec_{T,0}-\Tc\|_{1\rightarrow 1} \leq \varepsilon,\]
where $\|\Tc\|_{1\rightarrow 1}=\min_\rho(\|\Tc(\rho)\|_1/\|\rho\|_1).$
Exact reachability is recovered by letting $\varepsilon=0$. 

\noindent
\begin{defin}{\bf (CPTP controllability)}
{\em (i)} The control scenario $(\{H_c(t)\},\rho_A; \rho_B)$ is {\em completely CPTP-controllable}
if any CPTP map ${\cal T}$ on $S$ is reachable at some time $T_\Ti$, and similarly for 
$\varepsilon$-reachability. 
{\em (ii)} $(\{H_c(t)\},\rho_A; \rho_B)$ is 
{\em extreme CPTP-controllable} or, respectively, {\em m-rank CPTP-controllable}, 
if any CPTP map ${\cal T}$ that is extreme or, respectively, of Kraus rank at most $m$, 
is reachable at some time $T_\Ti$. 
\end{defin}

In the following, a key assumption will be that the control scenario is sufficiently powerful to allow for 
{\em complete unitary controllability of the joint system $SE$}; 
namely, we shall assume that the set of joint unitary operators that may be obtained by varying the  
control functions $u_\ell(t)$ and the corresponding control Hamiltonians $H_{SA,\ell}$ -- that is, 
of the form $U_{SE}(T)={\mathbb T}\exp\!\left[{-i\int_0^T H(t)dt}\right]$, 
where ${\mathbb T}\exp$ denotes the time-ordered exponential --   
is \emph{dense} in the full unitary group ${\mathfrak U}(\Hi_{SE}).$
While this may seem a very strong assumption, it is natural in our context for a number of reasons:

\begin{itemize}
\item By standard results in geometric control on Lie groups, at least in the case where $\Hi_E$ is effectively 
finite-dimensional (due, for instance, to the presence of an upper cutoff on energy), this is equivalent to require that the 
Lie algebra generated by the uncontrolled Hamiltonian, $-i H_0$, and the set of control Hamiltonians 
$\{-i H_{\ell} \otimes I_B\}$ in Eq. (\ref{eq:Hc}), is the full Lie algebra 
${\mathfrak u}(\Hi_{SE})$ of ${\mathfrak U}(\Hi_{SE})$.
It can be shown (see e.g. \cite{dalessandro-book,altafini-tutorial}) 
that almost every choice of just one pair of Hamiltonians on a single system will guarantee unitary controllability, 
and one may argue that a similar argument carries over to our composite-system setting as well. 
Hence, with {\em generic} choices of the Hamiltonians, joint unitary controllability will be guaranteed. 

\item In a series of papers, D'Alesssandro and collaborators \cite{albertini-indirect,dalessandro-indirect,dalessandro-indirecttwoqubit} studied the related concept of {\em indirect (unitary) controllability}. The control scenario is equivalent to ours, once the uncontrollable bath $B$ is removed, and unitary controllability is granted for the auxiliary system $A$: the task in this case is to determine under which condition all unitaries (and not all CPTP maps, as in our case) may be obtained by a joint evolution on $\Hi_{SA}$ followed by a partial trace on $A$. In \cite{dalessandro-indirect}, the authors show that, if the auxiliary system starts in the completely mixed state, $\rho_A=I/d_A,$ 
{\em joint unitary controllability is necessary} for indirect unitary controllability on $S$.  In particular, joint unitary controllability is 
necessary if requires wants {\em strong} indirect unitary controllability, namely, indirect unitary controllability {\em for all} initial states $\rho_A$ of the auxiliary system.
\end{itemize}  

\section{A subsystem principle for unitary design of CPTP maps}\label{sec:controllability}


\subsection{The standard approach: Stinespring dilation using a pure ancilla}

Consider an OSR of a CPTP map $\Ec$ as in Eq. \eqref{eq:kraus},  
and assume that the system Hilbert space $\Hi_S$ is paired to an auxiliary one, $\Hi_A$, of dimension $m$. 
One can the define a map $V:\Hi_S\rightarrow\Hi_A\otimes\Hi_S$ as 
\beq 
V \equiv \sum_{k=1}^m \ket{k}\otimes M_k.
\eeq
From the TP property, it follows that $V^\dag V=I_S$, thus $V$ is an isometric embedding. This is the dual of the {\em Stinespring representation} of a CP and unital $\Ec^\dag,$ and it is easy to see that it can be completed to a full unitary dilation $U_\Ec$ of $\Ec$ on $\Hi_A\otimes\Hi_S$. One way of achieving this consists in picking a reference state, say, $\ket{1},$ on $\Hi_A$ and identifying the action of $U_\Ec$ on the subspace spanned by $\ket{1}\otimes\ket{\psi_S}$ with the action of $V$ on $\ket{\psi_S}$:
\[U_\Ec\ket{1}\otimes \ket{\psi_S}=\Big(\sum_{k=1}^m \ket{k}\bra{1}\otimes M_k\Big)\ket{1}\otimes\ket{\psi_S}=V\ket{\psi_S}.\]
Since, as we already noted, $V^\dag V=I_S,$ we only need to choose the rest of $U_{\Ec}$ in such a way 
that $U_{\Ec}^\dag U_\Ec =I_{SA}.$
In the matrix block form induced by the tensor (Kroneker) product, with respect to the basis 
$\{\ket{k}\}$ of $\Hi_A$, this is equivalent to specify the first column of blocks of $U_\Ec$ as:
\beq\label{eq:Ublock}
U_{\Ec}=
\left[
\begin{array}{c|c|c|c}
 M_1 & *  & \ldots &*  \\
 \hline M_2 &  * &  \ldots &* \\
 \hline \vdots & *  & \ldots &*\\
 \hline M_m & * & \ldots &*  
\end{array}
\right],
\eeq
and then to complete the $*$ blocks by choosing a set of orthogonal columns for the full matrix. 
That the first $d_S$ columns are orthogonal follows by:
\[V^\dag V=\left[
\begin{array}{c|c|c|c}
 M_1^\dag & M_2^\dag  & \ldots & M_k ^\dag 
 \end{array}
\right]
\left[
\begin{array}{c}
 M_1   \\
 \hline M_2  \\
 \hline \vdots \\
 \hline M_m  
\end{array}
\right]=I_S. \]
By construction, we have:
\beq
\label{eq:SD}
\Ec(\rho)=\trace_A(U_\Ec \rho\otimes\ket{1}\bra{1} U_\Ec^\dag), \quad \forall \rho \in {\cal D}(\Hi_S).
\eeq
In particular, this construction proves that {\em any} CPTP map of rank $m$ can be obtained from 
an open-system evolution as in Eq. \eqref{eq:reduced}, provided that (i) the auxiliary system 
dimension is greater or equal than its Kraus rank, $d_A \geq m$, and (ii) its initial state $\rho_A$ is 
pure (so-called unitary representation theorem \cite{nielsen-chuang}).

\subsection{CPTP controllability results from virtual subsystems}

In the general setting we consider, where the environment $E$ comprises both $A$ and an uncontrollable 
bath $B$ in state $\rho_B$, and control over the auxiliary state $\rho_A$ may be limited, 
the question remains as to whether we can still engineer any desired CPTP map.
The key quantum resource is the ability to access a sufficiently pure ``portion'' of the environment, as captured 
by the general notion of a ``virtual subsystem'' \cite{viola-generalnoise,ZanardiVirtual}. 
A {\em virtual quantum subsystem}, say, $M$, of a larger system $E$ (the environment in our case) is 
associated with a tensor factor ${\Hi}_{M}$ of a {\em subspace} of $\Hi_E$:
\beq 
\Hi_E = ({\Hi}_{{M}} \otimes\Hi_F)\oplus \Hi_R,
\label{eq:subs}
\eeq
\noindent
for some factor ${\cal H}_F$ and possibly a remainder
space ${\cal H}_R$. The system $E$ is said to be {\em initialized in a virtual subsystem
${M}$ with state $\rho_M$} if its state may be decomposed as
$\rho_E= {\rho}_{M} \otimes\rho_F\oplus 0_R,$ where $0_R$ is the
zero operator on $\Hi_R$ and $\rho_F$ an arbitrary state on $\Hi_F$; in particular, 
following \cite{ticozzi-isometries,ticozzi-cooling}, two types of subsystem-initialization 
will play a key role in the present context:

\begin{defin}{\bf (Virtual-subsystem initialization)}
System $E$ is {\em initialized in a pure state of $M$} if 
$\rho_E = \ket{{\varphi}}\bra{{\varphi}}\otimes\rho_F\oplus 0_R,$
for some pure state $\ket{{\varphi}} \in {\Hi}_{M}$. 
Similarly, $E$ is {\em $\varepsilon$-approximately initialized in a pure state of $M$} if 
there exists a pure-state initialization of $E$, $\tilde{\rho}_E = 
\ket{\tilde{\varphi}}\bra{\tilde{\varphi}} \otimes \rho_F\oplus 0_R$, such that \beq
\label{condpu} 
d_{\rm TV}(\rho_E,\tilde{\rho}_E) \leq \varepsilon. 
\eeq
\end{defin}

The following is a central result of the paper, effectively deriving sufficient conditions for the design of a map 
with a given Kraus rank from a ``subsystem principle'':

\begin{thm}[{\em m}-rank CPTP controllability]
\label{thm1}  Assume joint unitary controllability and factorized initial conditions 
$\rho_S \otimes \rho_E$.  Then the target system $S$ is {\em $\varepsilon$-approximate m-rank CPTP 
controllable}  if there exists a decomposition 
of $\Hi_E$ as $\Hi_E = ({\Hi}_{{M}} \otimes\Hi_F)\oplus \Hi_R$, with dim$\,(\Hi_M)=m$, 
such that $\rho_E$ is $\varepsilon$-approximately initialized in a pure state of $M.$ \end{thm}
\proof
We will show that, under the hypothesis, every $\Tc$ on $S$ of Kraus rank $m$ or less  is {$\varepsilon$-reachable} via a generalized Stinespring construction.
If $\rho_E$ satisfies the condition in Eq. \eqref{condpu}, then we may write 
\beq
\rho_E \equiv \tilde{\rho}_E +\Delta\rho_E,\quad
\frac{1}{2}\trace(|\Delta\rho_E|)\leq\varepsilon.
\label{eq:deviation}
\eeq
If $\varepsilon=0$, we can use the Stinespring construction described above and engineer $\Tc$ perfectly, 
by defining a unitary 
$U_\Ti\in{\mathfrak U}(\Hi_S\otimes\Hi_M)$ as in Eq. \eqref{eq:Ublock}, and then extending its action to the whole 
$\Hi_{SE} = (\Hi_S\otimes \Hi_M \otimes \Hi_F)\oplus(\Hi_S\otimes \Hi_R)$ as 
\( W_{SE} \equiv (U_{\Ti}\otimes I_F)\oplus  I_{SR} .\)

Next, for $\varepsilon >0$, 
we show that by applying the same unitary $U_\Ti$ to $\rho_S\otimes\rho_E,$ even when $\rho_E$ is only 
$\varepsilon$-approximately initialized in a pure state, the resulting state is $\varepsilon$-close to the target output 
for all initial states. In fact, we have:
\begin{eqnarray*}
\Ec(\rho_S) & = & \trace_E(W_{SE}\,\rho_S\otimes\rho_E\, W_{SE}^\dag) \nonumber \\
&= & \Tc(\rho_S)+\trace_E [W_{SE}\,\rho_S\otimes\Delta\rho_E \,W_{SE}^\dag] \nonumber \\
&\equiv &\Tc(\rho_S) + \tilde{\cal E}(\rho_S \otimes \Delta\rho_E),
\end{eqnarray*}
where $\tilde {\cal E}$ is a TP CP map and hence a trace-norm contraction \cite{nielsen-chuang}.  
Then, from Eq. (\ref{eq:deviation}), it follows that
\( d_{\rm TV}(\Ec(\rho_S),  \Tc(\rho_S))\leq\varepsilon,\) for all $\rho_S \in {\mathcal D}(\Hi_S)$. 
Hence the same $W_{SE}$ also ensures $\varepsilon$-approximate engineering of $\Tc.$
\qed

\noindent
It is interesting to note that the standard Stinespring dilation of Eq. (\ref{eq:SD}) is recovered as a special case of the above 
controllability result in the exact setting, $\varepsilon=0$, by letting $m=d_S^2,$ $\Hi_F= {\mathbb C},$ and $\Hi_R=\emptyset$.

\vspace*{1mm}

{\em Remark:} The existence of a subsystem of $\Hi_E$ of dimension $m$ that is 
$\varepsilon$-approximately initialized in a pure state is equivalent to the existence of a subspace, 
say, $\Hi_1 \leq \Hi_E,$ {\em of dimension $d_1 \leq d_E/m$}, 
such that $\rho_E$ restricted to $\Hi_1$ has trace equal to 
at least $(1-\varepsilon).$ That the initialization implies the existence of such a subspace is clear by considering 
$\Hi_1 \equiv \span\{\ket{\varphi}\}\otimes \Hi_F,$ and the converse implication follows by the same identification, 
completed by defining additional $(m-1)$ subspaces $\Hi_j \simeq \Hi_1$, $j=2,\ldots , d_S,$ 
and identifying them with $\Hi_j \equiv \span\{\ket{\varphi_j}\}\otimes \Hi_F,$ 
where the $\ket{\varphi_j}$ complete $\ket{\varphi}$ to an orthonormal basis for $\Hi_M.$
All the examples examined in Ref. \cite{ticozzi-cooling}, in which the above construction is carried out explicitly, 
also work in the present setting. In particular, we know how to construct  a $\varepsilon$-pure subsystem in 
thermal environments and in $n$-qubit environments, under certain constraints on the entropy.
However, it is also clear that having a pure subsystem is, in general, {\em not} necessary for CPTP controllability, 
as we will show explicitly in Sec. \ref{sec:maj}.

\vspace{1mm} 

By combining the previous theorem with the characterization of exact purification and cooling 
obtained in Ref. \cite{ticozzi-cooling}, we also have the following:
 
\begin{prop}[Extreme CPTP controllability]
\label{prop:extremecont}
The target system is {\em extreme CPTP-controllable} {if and only if} there exists a decomposition 
$\Hi_E = ({\Hi}_{{M}} \otimes\Hi_F)\oplus \Hi_R$, with dim$\,(\Hi_M)=d_S$, and $\rho_E$ initialized in 
a pure state of $\Hi_M.$
Furthermore, the system is {\em $\varepsilon$-approximately extreme CPTP-controllable} if there exists a decomposition of $\Hi_E$ as above, 
and $\rho_E$ is $\varepsilon$-approximately initialized in a pure state of $\Hi_M.$
\end{prop}

\proof
Since extreme maps must have a Kraus representation with operators satisfying Eq. \eqref{eq:extreme}, their Kraus rank can be at most $d_S$. Thus, the existence of a pure initialization of a $d_S$-dimensional subsystem is sufficient for their reachability given Theorem \ref{thm1}. Necessity follows from the fact that any map that has a single pure 
state as output is extreme, and the main theorem of \cite{ticozzi-cooling} shows 
that the existence of a $d_S$-dimensional pure subsystem of $E$ is a necessary condition to attain these maps.

The (sufficient) conditions for $\varepsilon$-approximate controllability, 
follows from a direct application of the same contraction argument used in the proof of Theorem \ref{thm1}.
\qed

{\em Remark 1:} Notice that, while a $d_S$-dimensional pure subsystem is required in order to be able to engineer 
{\em any} extreme map, there are some maps for which this is clearly \emph{not} necessary. For example, unitaries are extreme 
maps, they have Kraus rank one and, under the joint unitary controllability assumption, they do not need any auxiliary resources to be enacted.

{\em Remark 2:} The above result poses a clear no-go to the possibility of {\em strong} CPTP controllability in the sense 
of \cite{dalessandro-indirect}, that is, under the requirement that the auxiliary subsystem state be arbitrary: if 
$d_E < \infty$, it is possible to find states of $E$ that do not admit a decomposition with a pure subsystem 
of dimension $d_S$ \cite{ticozzi-cooling}. By the above Proposition, this prevents reachability for some extreme maps, 
and hence CPTP-controllability with an arbitrary environment state.

\subsection{Probabilistic unitary design} 
\label{sec:prob}

Theorem \ref{thm1} shows, by extending Stinespring's construction, that CPTP controllability is certainly guaranteed 
if there exists a $d_S^2$-dimensional subsystem of the environment which is initialized in a pure state, and we 
have joint unitary controllability.  However, this is only a sufficient criterion, and 
CPTP controllability may still be possible with less taxing resources, for instance, a smaller ancilla.

In this regard, Lloyd conjectured in Ref. \cite{lloyd-science} that an ancillary system of minimum dimension $d_S$ would 
suffice to ensure complete CPTP-controllability, provided one could initialize it in any state, pure or mixed. However, 
this conjecture has been proven wrong in Refs. \cite{terhal,other,other1}: one may explicitly identify 
CPTP maps that need a larger (at least $d_S+1$ dimensional) ancilla in order to be implemented 
via unitary design as in Eq. \eqref{eq:reduced}. This also proves that the condition in Proposition \ref{prop:extremecont} 
is {\em not sufficient} to have complete CPTP controllability.  While in fact the work by Lloyd \& Viola in Ref. \cite{Lloyd2002} 
shows that an ancilla $A$ as small as a single pure qubit does suffice provided it is {\em resettable} (see also Sec. 
\ref{subsec:lv}), from a subsystem-principle perspective this in any case implies the existence of a pure qubit subsystem 
in the ``physical'' environments $E$ needed to purify $A$ on each use.  Alternatively, remaining within the coherent 
Hamiltonian setting under consideration, we may relax the requirement that the target map ${\cal T}$ is implemented
deterministically in a ``single shot'': as we now show, by allowing for some \emph{classical} 
resources and a \emph{non-deterministic} construction, (exact) CPTP controllability is indeed regained, 
if we have access to a pure, $d_S$-dimensional ancilla subsystem, as Lloyd originally conjectured.

In order to formalize the idea, we extend the unitary design method to include classical stochastic resources, 
which can be used to implement mixtures of evolutions, and hence simulate, on average, CPTP maps that are 
not extreme \footnote{The idea of ``average realization'' of CPTP dynamics has been used previously, 
e.g. in Ref. \cite{albertini-feedback,Piani2011}. However, the first work addresses only {\em state} controllability via 
measurement-based feedback; in the second paper, which is specifically tailored to optical 
qudit channels, the implementation is non-deterministic in the sense that {\em only a finite probability of success} 
can be achieved in general.}.

\begin{defin}{\bf (CPTP reachability on average)} 
Given a probability distribution $\pi\equiv \{\pi_j\}$, a CPTP map ${\cal T}$ is reachable {\em on average} at 
time $T$ if there exists joint unitaries  $U_{SE,j}(T)$ and each can be enacted with probability $\pi_j$ so that
\[ \Ti(\rho_S)=\mathbb{E}_\pi\{\trace_E  [U_{SE,j}(T)\,\rho_S\otimes\rho_E \, U_{SE,j}^\dag(T)]\}{ .}\]
We say that the target system is {\em CPTP-controllable on average} if every CPTP map is reachable on average 
as above.
\end{defin}

\noindent We can then establish the following:

\begin{thm}[Probabilistic reachability on average]
Let ${\cal T}$ be an arbitrary CPTP map on a $d_S$-dimensional system, and assume that we can sample from any 
{\em classical} distribution $\pi\equiv \{\pi_j\}$ on $d_S^4$ elements.  ${\cal T}$ is reachable on average at time 
$T$ if there exists a decomposition $\Hi_E = ({\Hi}_{{M}} \otimes\Hi_F)\oplus \Hi_R$, 
with dim$\,(\Hi_M)=d_S$ and $\rho_E$ initialized in a pure state of $\Hi_M$. The system is then CPTP-controllable on average.
\end{thm}

\proof
Proposition \ref{prop:extremecont} guarantees that we can implement exactly, without the need of classical 
randomization, all extreme maps. Any other map $\Ti$ can be written as $\Ti=\sum_j\pi_j \Ti_j,$ where $\Ti_j$ is 
extreme. Any $\Ti$ can be parametrized (for example, by using Choi's matrix) 
as a convex set immersed in a $d_S^2 \times d_S^2$-dimensional space of complex positive-semidefinite matrices, 
and these in turn can be re-parametrized as a $d_S^4-1$ {\em real} vector. By Carath\'eodory's theorem on the convex 
hull \cite{caratheodory}, there are at most $d_S^4$ components in the sum, and the result follows.
\qed

{\em Remark 1:}  While we have distinguished this way to simulate the output of a target CPTP map from the one described in Eq. \eqref{eq:reduced} as {\em probabilistic}, the actual difference is subtle: distinguishing the outputs would be possible via measurement statistics only if we had multiple identical copies that use the same classical stochastic resource. 
Furthermore, it is instructive to think about maps that can output pure states: it is immediate to see that, for these maps, every map in an equivalent convex combination should also output that same pure state to the corresponding input, otherwise the convex combination would not. This indicates that in the probabilistic channel-design approach, 
the classical resources are used only to simulate the {\em classical uncertainty} in the description of the target map,
encoded in the associated convex weights.

{\em Remark 2:} $\varepsilon$-approximate average CPTP-map engineering can be also guaranteed, 
in the same spirit of unitary engineering, by requesting a $\varepsilon$-pure state in the virtual subsystem. 
In this way, each extreme map entering the decomposition of ${\cal T}$ can be obtained within 
$\varepsilon$-precision, and the average error will be upper-bounded by $\sum_j\pi_j\varepsilon=\varepsilon.$

\section{Unitary design of CPTP maps beyond the subsystem principle}
\label{sec:maj}

\subsection{A majorization approach for stochastic unitary maps}
\label{sec:stochuni}
In the previous sections, we have focused on deriving a subsystem principle for CPTP controllability maps using a Stinespring-type construction: 
{access to a virtual subsystem of the environment initialized in a $(\varepsilon$-)pure state suffices for all possible target maps to be 
reached. However, if the task of interest is to engineering a specific map ${\cal T}$ or a set of maps that does not include extreme ones,} 
this need {\em not} be necessary.

The key assumption is that the spectrum of the environment state $\rho_E$ majorizes the set of convex 
weights needed to write $\Ti$ as a convex combination of other extreme maps.
Recall that a probability distribution $\{p_j\}$ is said to {\em majorize} another distribution on 
the same set $\{q_j\}$ if the following conditions hold:
\[\sum_{j=1}^k p_j\geq \sum_{j=1}^k q_j,  \quad \forall k\geq 1,  \]
in which case we write $\{p_j\}\succeq \{q_j\}.$
It is a well-known result \cite{horn-johnson} that 
$\{p_j\}\succeq \{q_j\}$ if and only if there exists a unitary $V$, such that $q_i=\sum_j |V_{ij}|^2 p_j$ or, 
compactly in matrix form, by defining $P_{ij} \equiv |V_{ij}|^2$, we may write \(\vec p= P \vec q.\) 
Any matrix $P$ that can be obtained as the element-wise modulus square of a unitary $V$ is called a 
{\em unistochastic (or ortho-stochastic)} matrix.
A well studied class of (unital) CPTP maps that are non-extreme comprises {\em stochastic unitaries}, 
$$\Ti(\rho)=\sum_{j=1}^m q_jU_j\rho U^\dag_j, \quad \sum_{j=1}^m q_j =1, \; q_j \neq 0,$$
where $U_j\in \mathfrak{U}(\Hi_{S}).$
We first show how it is possible to reach this class of maps using majorization, {\em without} constraining the purity 
of a virtual-subsystem initialization:

\begin{thm}[Reachability of stochastic unitaries] 
Assume joint unitary controllability controllability and factorized initial conditions 
$\rho_S \otimes \rho_E$ on $\Hi_S \otimes \Hi_E$.  Let $\Tc$ be stochastic unitary, with weights $\{q_j\}$. 
Then, for every $\varepsilon \geq 0$, $\Tc$ is {\em $\varepsilon$-approximate reachable} if there exists a decomposition of $\Hi_E$ 
as $\Hi_E = ({\Hi}_{{M}} \otimes\Hi_F)\oplus \Hi_R$, with dim$\,(\Hi_M)=m$,
and an initialization of $E$ in $M$, $\tilde{\rho}_E =
{\rho_M} \otimes \rho_F\oplus 0_R$, such that:
\begin{enumerate}
\item $\rho_M=\sum_{j=1}^m p_j\ket{j}\bra{j},$ with $\{p_j\}\succeq\{q_j\}$;
\item $
d_{\rm TV}(\rho_E,\tilde{\rho}_E) \leq \varepsilon. $
\end{enumerate}
\end{thm}
\proof
With respect to the decomposition $\Hi_E = ({\Hi}_{{M}} \otimes\Hi_F)\oplus \Hi_R$, 
define a joint unitary of the form
\[W_{SE}\equiv C_U(I_S\otimes V_E),\]
where
\beq
C_U=\sum_j \left( U_j\otimes\ket{j}\bra{j}\right)\otimes I_F \oplus \left(I_{S}\otimes I_ R\right),
\label{eq:CU}
\eeq
implements $U_j$ conditionally on the state of $M$, and $V_E=V\otimes I_F\oplus I_R$ is a unitary, 
with $V$ such that $q_i=\sum_j |V_{ij}|^2 p_j$.  Note that such a $V$ exists given hypothesis (i).  
Then in the case $\varepsilon=0$, by noticing that ${\rm diag}(V\rho_M V^\dag)=(q_1,\ldots,q_m),$ 
the above $W_{SE}$ is such in Eq. \eqref{eq:reduced} $\Ec_{T,0}=\Tc$ by construction.
The fact that $W_{SE}$ also works when $\varepsilon\neq 0$ follows from the same contraction 
argument used in proving Theorem \ref{thm1}.
\qed 

\subsection{Sufficient conditions for unitary design of general convex-combination maps}
\label{sec:blockconvex}

A similar majorization-based approach can be used to derive sufficient conditions for the unitary design of more general 
convex combination of CPTP maps. For the sake of simplicity, we exemplify the construction for the binary case:
\[\Ti=q_1\Ti_1 + q_2\Ti_2,  \quad q_1+q_2 =1 .\]
Since the maps $\Ti_j$ are extreme, by Proposition \ref{prop:extremecont} they may be obtained as in Eq. \eqref{eq:reduced} 
if and only if the dimension of the corresponding ancilla $m=d_S.$ Let $U_{\Ti_j}$ denote the unitaries that implement $\Ti_j$ on 
$\Hi_S\otimes\Hi_{M,j}$, as in the proof of Theorem \ref{thm1}. 

Assume that there exist subspaces of $\Hi_E,$ say, $\Hi_1\equiv \Hi_{M_1}\otimes\Hi_{F,1}$ and $\Hi_2\equiv \Hi_{M_2}\otimes\Hi_{F,2}$, with ${\rm dim}(\Hi_{M_j})=d_S$, such that $\rho_E$ restricted to each of them has the form $\rho_E|_{\Hi_j}=\ket{j}\bra{j}\otimes\tilde\tau_j$, where $\trace({\tilde\tau_j})=q_j,$ respectively. Then we can build a decomposition of the environment Hilbert space as:
\beqan\Hi_{E}&=&\Hi_1\oplus\Hi_2\oplus\Hi_R\\
&=&(\Hi_{M,1}\otimes\Hi_{F,1})\oplus(\Hi_{M,2}\otimes\Hi_{F,2})\oplus\Hi_R,
\eeqan
and, including $\Hi_S$, we have, accordingly:
\beqan&& \Hi_S\otimes \left[ (\Hi_{M,1}\otimes\Hi_{F,1})\oplus(\Hi_{M,2}\otimes\Hi_{F,2})\oplus\Hi_R\right]=\\
&&(\Hi_S\otimes\Hi_{M,1}\otimes\Hi_{F,1})\oplus(\Hi_S\otimes\Hi_{M,2}\otimes\Hi_{F,2})\oplus(\Hi_S\otimes\Hi_R).\eeqan
In analogy to Eq. (\ref{eq:CU}), and relative to the three orthogonal subspaces in the above decomposition, let:
\[{C}_{SE}^{(2)} \equiv (U_{\Ti_1}\otimes I_{F,1})\oplus (U_{\Ti_2}\otimes I_{F,2})\oplus (I_{S} \otimes I_R).\]
It can be verified by direct computation that if, as we assumed, $\rho_E$ allocates probability $q$ on the subspace $\Hi_{M_1}\otimes \Hi_{F_1},$ and $(1-q)$ on $\Hi_{M_2}\otimes \Hi_{F_2},$ then $C_{SE}^{(2)}$ will implement $\Ti$ as in Eq. \eqref{eq:reduced}.
This construction can be directly extended to an arbitrary number $K>2$ of extreme maps, by identifying more subspaces $\Hi_i$ of 
dimension multiple of $d_S$, that account for the correct amount of probability, and by letting
\[{C}_{SE}^{(K)} \equiv \bigoplus_{k=1}^K(U_{\Ti_k}\otimes I_{F,k})\oplus (I_{S} \otimes I_R).\] 

The existence of the required $\Hi_j$ subspaces can be checked via the following algorithm, which constructs, if possible, 
a choice of subspaces associated with a probability distribution that majorizes the $q_k$:
\begin{itemize}
\item[(I)]
Diagonalize $\rho_E$ and order the basis so that the eigenvalues $\lambda_\ell(\rho_E)$ are non-increasing in 
$\ell$. Order the convex weight set $q_k$ accordingly.

\item[(II)] Define $f$ to be the smallest number so that the first $f$ eigenvalues of $\rho_E$ are larger than the sum of the first $f$ elements $q_\ell$. Check if the following holds for every $k$:
\beq
\sum_{\ell=1}^{kf}\lambda_\ell(\rho_E)\geq \sum_{\ell=1}^k q_\ell . 
\label{eq:majsum} 
\eeq 
If this is not the case, this method is not viable for engineering the target convex combination.

\item[(III)] Assuming that Eq. (\ref{eq:majsum}) holds, note that 
the remaining ``degrees of freedom'' in $\Hi_E$ must be at least $(d_S-1)Kf$ and, by construction, they all belong to 
the kernel of $\rho_E.$ If this is the case, we can identify:
\[\span\{{\ket{1_\ell}}\}\otimes\Hi_{F,\ell} \equiv \span\{\ket{\lambda_\ell(\rho_E)},\ell=1,\ldots,f\},\] 
and complete each subspace to $\Hi_{M,\ell}\otimes\Hi_{F,\ell}$ by adding elements of the kernel. 
By construction, all $\Hi_{F,\ell}\simeq \Hi_F$ are isomorphic and of dimension $f.$
With respect to this decomposition, the state of the environment takes the form:
\[\rho_E=\bigoplus_\ell \tilde q_\ell\ket{1_\ell}\bra{1_\ell}\otimes \tilde{\tau}_\ell,\]
where $\tilde{\tau}_\ell$ are density operators on $\Hi_{F,\ell}$ and $\{\tilde q_\ell\}$ a probability distribution that majorizes $\{q_\ell\}.$ 

\item[(IV)] Let $V$ be a unitary such that $q_j=\sum_\ell |V_{ j\ell}|^2\tilde q_\ell.$
The desired non-extreme map ${\cal T} =\sum_j^K q_j {\cal T}_j$ can then be engineered by 
a block-unitary $C_{SE}^{K)}$ as above, after the action of 
$V\otimes I_F$ on $(\bigoplus_{\ell}\span\{{\ket{1_\ell}}\})\otimes\Hi_{F,\ell}.$
\end{itemize}

\vspace{1mm}

{\em Remark:} For a general non-extreme map, that obeys Eq. \eqref{eq:majsum} but may 
involve up to $d_S^4$ extreme maps in its convex-sum decomposition,
the construction we presented requires an environment whose dimension is 
of the order of $d_S^5$. While this is clearly much more demanding in terms of unitary control than the 
standard Stinespring construction, it can be justifiable in limiting cases, in particular in situations where 
no access to a (nearly) pure auxiliary state $\rho_A$ is granted.

\section{Toward optimal CPTP design via channel-state duality}
\label{sec:optimal}

The previous controllability results rely on the ability of enacting certain unitary evolutions on the joint space of the system of interest 
and its  environment, guaranteed by joint unitary controllability. Checking if this assumption holds for a given control scenario is relatively straightforward, and computationally tractable -- upon constructing the control Lie algebra generated by the available Hamiltonians and comparing it to the full one. However, how to explicitly synthesize a control that achieves the intended evolution is, 
in general (and already at the closed-system, unitary level), a much harder problem. In this section, we propose a way to recast the 
CPTP control synthesis problem based on unitary design as an optimization problem. The target map is going to be reached exactly whenever the {\em cost} to minimize reaches zero. While the problem is guaranteed to have a solution with zero cost if the joint system is unitary controllable, this reformulation can also be useful to investigate whether exact reachability is possible without full joint controllability, or to probe the actual accuracy of the control synthesis for approximate engineering in the presence of mixed virtual subsystem states.

Instead of writing the problem directly for the target CPTP map, the idea is to formulate an {\em equivalent state-to-state transfer problem}, albeit on a larger state space, via the Choi-Jamio{\l}kowski representation. One advantage in doing this is that, in principle, a wide array of algorithms are available for {optimal state-transfer control problem} \cite{dalessandro-book,qoptimal-comparison}.  
Care is needed, nonetheless, since the resulting problem is an atypical one, from the standpoint of standard optimal control algorithms: 
while most of the optimal state transfer literature considers a ``full'' state transfer, in our setting the 
desired output is specified {\em only on a subsystem.} 
 
Formally, as we mentioned in Sec. \ref{sec:prel}, in order to exploit state-channel duality 
we couple the joint system and environment Hilbert space to an isomorphic copy of $\Hi_S,$ which 
we call $\Hi_{S'}.$ On $\Hi_{S'}\otimes \Hi_S\otimes \Hi_E$, 
we consider the initial state $\Phi\otimes\rho_E,$ where $\Phi$ is the maximally entangles state, as in  
Eq. \eqref{eq:choi}. Next, we let a joint unitary $U_{SE}$ act on $\Hi_S\otimes\Hi_E$, while the trivial (identity) evolution 
is enacted on $\Hi_{S'},$ and we take a partial trace over $\Hi_E.$ 
Let us denote by $\Phi'$ the resulting density operator on $\Hi_{S'}\otimes\Hi_S.$

\begin{figure}[t]
\hspace{-3mm}
\centering
\includegraphics[width=0.45\columnwidth]{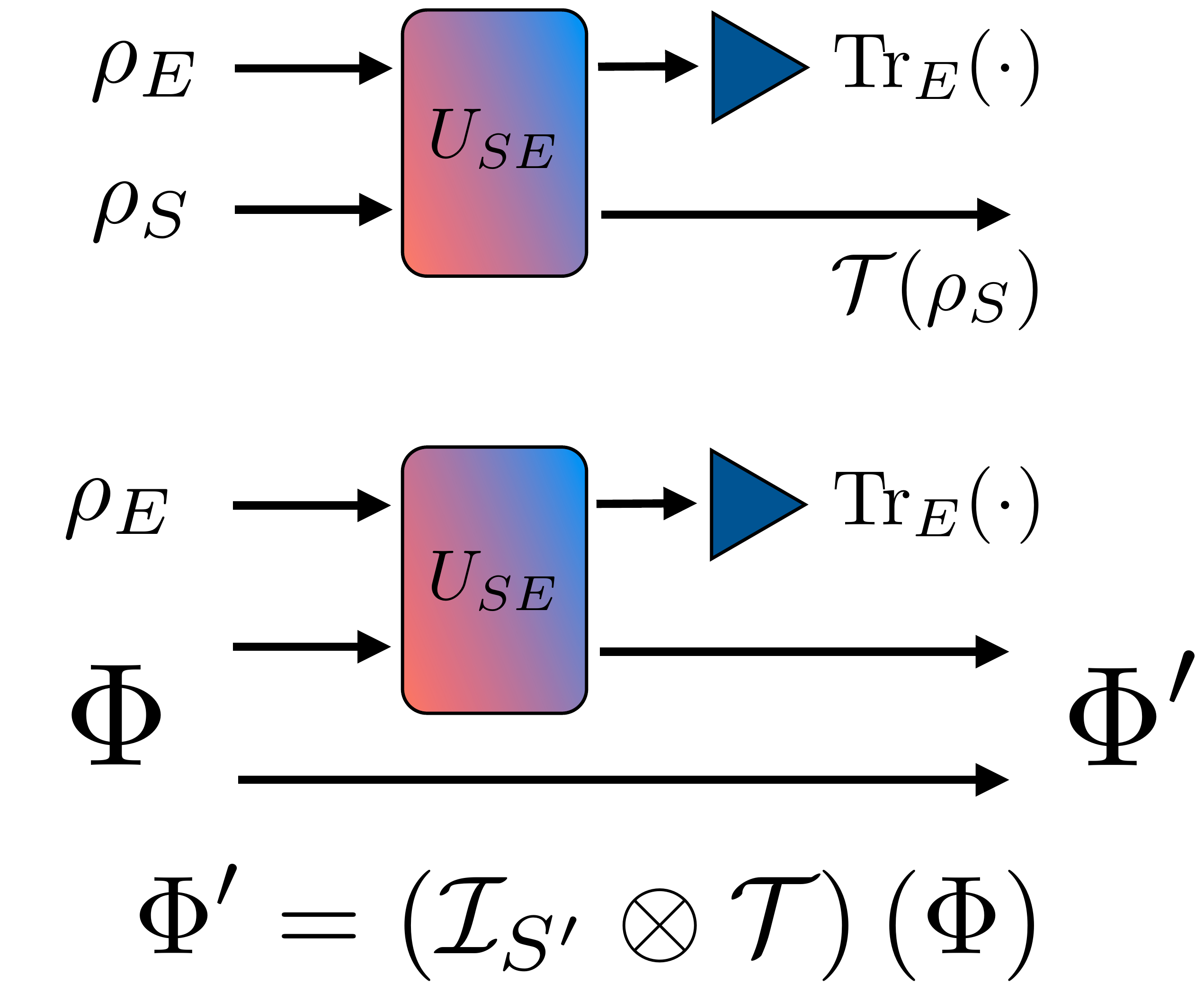}
\vspace*{-1mm}
\caption{(Color online) Remapping the unitary design of a target CPTP map ${\cal T}$
(circuit on the top part of the image) to unitary transfer of a single state (circuit on the bottom) 
via the Choi-Jamio{\l}kowski isomorphism. } 
\label{fig:optimal}
\end{figure}

As a consequence of the Choi-Jamio{\l}kowski isomorphism, namely, the one-to-one correspondence between CPTP maps and Choi matrices, it follows that $U_{SE}(T) $ enacts a target map $\Ti=\Ec_{T,0}$ as in Eq. \eqref{eq:reduced} {\em if and only if} 
$\Phi'= ({\cal I}_{S'}\otimes \Ti ) (\Phi)$ (see also Fig. \ref{fig:optimal} for a pictorial rendering).
We can then use this equivalence to recast the ``dissipative gate synthesis'' problem associated to ${\cal T}$
into an optimal state transfer problem from $\Phi \mapsto \Phi'$. 
Let $H_0,H_\ell\equiv H_{SA,\ell}$ be the free and control Hamiltonians as introduced in 
Sec. \ref{sec:prel}, and let $\{\sigma_i\}$ be a basis for the operator space ${\cal B}(\Hi_{S'S}).$ Define the components:
\[c_i \equiv \trace[\sigma_i \,({\cal I}_{S'}\otimes \Ti)(\Phi)].\]
We can then look for the best choice of unitary $U_{SE}$ that implements $\Ti$ at time $T$ by posing an optimization 
problem of the following type:
\begin{eqnarray*}
&{\textrm{minimize}}_{\:U_{SE}} & \sum_i \left|c_i- \trace[\sigma_i\otimes I_E 
[(I_{S'}\otimes U_{SE})\Phi\otimes\rho_E(I_{S'}\otimes U_{SE}^\dag)] \right|^2 , \\
& \textrm{subject to\ }& 
U_{SE}(T) ={\mathbb T}\exp\bigg\{\!-i\!\int_0^T\Big[H_0+\sum_\ell u_\ell(t) H_\ell \Big] dt\bigg\}.
\end{eqnarray*}
In fact, once $H_0$ and $\{H_\ell\}$ are specified, 
the actual optimization variables are the control inputs $u_\ell(t)$ in Eq. (\ref{eq:Hc}): in the above form, 
the problem is written as if the $U_{SE}$ was the variable for the sake of compactness.
Note that here the cost function is the quadratic distance between the components of the desired state on 
$SS'$ and the one corresponding to the chosen controls, but other choices may be better suited to the scope \cite{grace2010}. 
The problem is {\em guaranteed} to have a solution for which the cost function is zero if the system of interest is CPTP controllable. 
Developing an explicit algorithm, or a suitable adaptation of an existing one, to solve the above type of optimal-control problems is an 
interesting direction for future investigation.

\section{Illustrative applications}
\label{sec:examples}

In this section, we revisit some simple existing protocols for engineering a target CPTP dynamics within the present 
framework of unitary design problems. We highlight their use of limited resources, both in terms of the dimension of 
the environment and in the types of available joint dynamics. Since we work within a coherent Hamiltonian-control 
setting, we stress that having access to a suitable set of \emph{conditional} operations, 
along with other (protocol-dependent) control resources, will be instrumental in order to replace classical 
(measurement-based) feedback protocols with protocols employing only coherent feedback.

\subsection{Coherent implementation of binary-tree protocols for channel construction}  
\label{subsec:lv}

The first method for universal approximate engineering of arbitrary CPTP maps was proposed in Ref. \cite{viola-engineering}. 
In terms of auxiliary quantum resources, it only needs the {\em smallest possible} quantum environment: a single qubit, 
albeit the latter must be resettable in a known pure state. While in practice the resulting map is obtained 
with accuracy $\varepsilon >0$, due to the presence of a Hamiltonian coherent averaging procedure \cite{viola-dd}, 
the accuracy is only limited by how fast the averaging cycle can be enacted. 
The original proposal relies crucially on discrete-time, measurement-based single-bit feedback, with an explicit 
``binary-tree'' construction being provided to implement the required generalized quantum measurement.  This 
construction has been subsequently improved both in terms of making contact with specific universal gate sets and 
in terms of efficiency \cite{ItenCircuits,yale}. 
Here, we show how to achieve the same task by using only coherent evolutions, 
at the cost of substituting the single resettable auxiliary qubit with multiple copies of the same, 
if higher-rank maps are considered. In the light of Proposition \ref{prop:extremecont}, we know that having access 
to a sufficiently large (albeit not necessarily pure) auxiliary subsystem is {\em unavoidable} if we aim to unitarily 
engineer a set of evolutions which contains {\em all} the extreme ones.

\subsubsection{Rank-two channels with pure ancilla.}
Having a pure ancilla of rank 2 guarantees exact CPTP controllability of the system via a
Stinespring-type dilation, yet does not provide a constructive procedure to synthesize 
an effective joint unitary evolution in terms of the available resources. The protocol we describe, 
on the other hand, provides a sequence of unitary evolutions that approximates an effective $U_{SE}(T),$ 
which we know exists, by using only a specific class of controlled operations. 
The relevant resources and task may be summarized as follows:

\begin{description}
\item[Task] To approximately enact an  arbitrary CPTP map 
${\cal T}$ of Kraus rank 2 on a $d_S$-dimensional target system $S$.

\item[Environment] A two-level system, with state space $\Hi_{E}\equiv \span\{\ket{0},\ket{1}\}$, initially in a pure state, say, 
$\rho_E=\ket{0}\bra{0}$. No interaction with an uncontrollable bath is assumed to be present.

\item[Control resources] We need an entangling Hamiltonian of the form $H_{SE} \equiv \gamma \Pi_S\otimes X_E,$ 
where  $\gamma >0$ is a tunable parameter and $\Pi_S=\ket{\phi}\bra{\phi}$ is a projector 
onto a pure state of $S$, while $X_E\equiv \ket{0}\bra{1}+\ket{1}\bra{0}$. In addition, 
complete (ideally, instantaneous) Hamiltonian control is required on $S$ in order to implement Hamiltonian averaging, 
as well as arbitrary conditional unitaries of the form:
\[U_0\otimes \ket{0}\bra{0}+U_1\otimes\ket{1}\bra{1},   \]
where $U_{0,1}$ are arbitrary unitaries on $S$ \footnote{In the original protocol \cite{viola-engineering}, 
these conditional unitaries are substituted by measurement-based feedback and unitary control on $S$ alone, 
conditional on the output of the feedback.}. 
\end{description}

Given the above resources, the construction is based on two simple mathematical observations: (i)
every CPTP map with Kraus rank 2 is associated to Kraus operators $M_0,M_1$ that admit a polar 
decomposition \cite{horn-johnson} of the form
\beq
\label{eq:polar} 
M_0=U_0\cos(t\gamma P),\quad  M_1=U_1\sin(t\gamma P),
\eeq
where $P$ is a positive-semidefinite operator on $S$. This follows from the fact that if $M_0,M_1$ correspond to a CPTP map, they satisfy $M_0^\dag M_0 +M_1^\dag M_1 =I,$ which implies that their respective polar components must be lesser or equal than the identity, simultaneously diagonalizable and their square sum to the identity. (ii) Owing to the spectral theorem, 
every positive-semidefinite $P$ can be written as $P=\sum_k\lambda_kV_k\Pi_SV_k^\dag,$ with $\lambda_k$ convex weights, $\sum_k \lambda_k=1$.
Based on these observations, we now show that it is possible to approximate (to arbitrary accuracy, 
in principle) any rank-two CPTP evolution by the following sequence of coherent dynamics 
(see also Fig. \ref{fig:lloydviola}):

\begin{figure}[t]
\hspace{14mm}
\includegraphics[width=0.9\columnwidth]{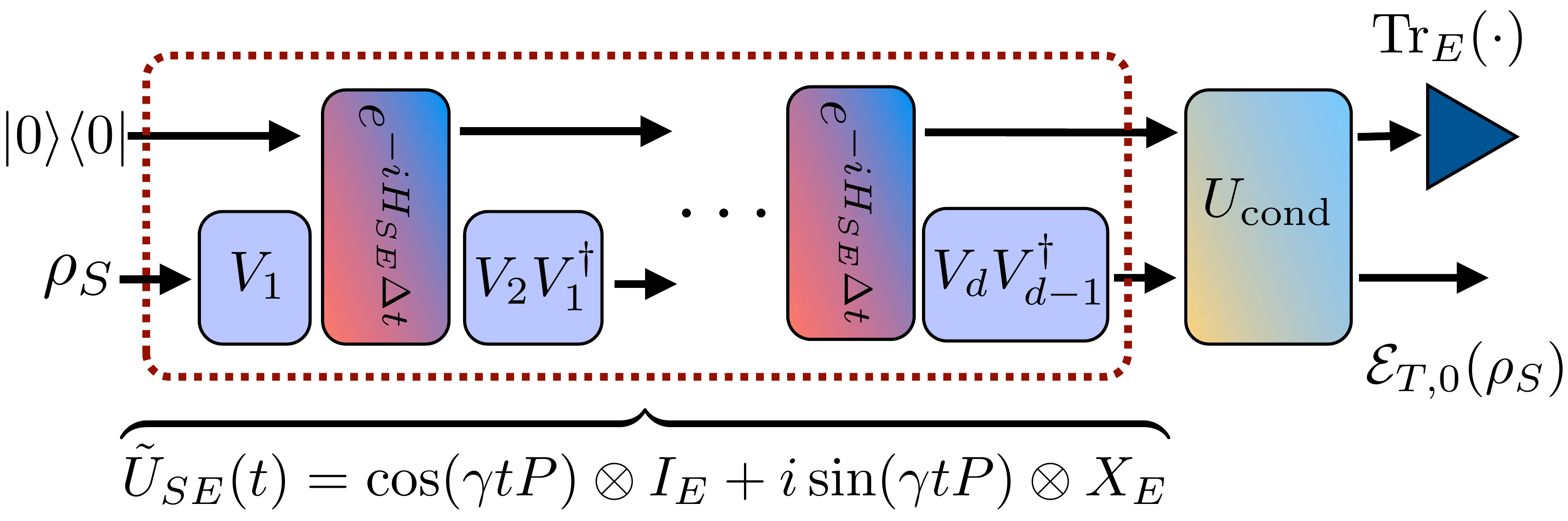}
\vspace*{-8mm}
\caption{(Color online) Coherent implementation of the Lloyd-Viola protocol \cite{viola-engineering} 
for approximate design of CPTP maps in the special case of a rank-two target map. } 
\label{fig:lloydviola}
\end{figure}

\begin{enumerate}
\item[(I)]  Assume that ${\cal T}$ has a OSR in terms of a pair of Kraus operators with a polar decomposition as in Eq. \eqref{eq:polar}. Let $S$ and $E$ be coupled by $H_{SE}$ as above, and evolve for a time interval $t,$ while applying an Hamiltonian averaging technique \cite{viola-dd,pawel2014} aimed to simulate $\gamma P\otimes X_E$, with 
$P$ an arbitrary positive semidefinite operator, as the new effective Hamiltonian. In view of the above observations, 
this is possible by dividing a finite evolution interval $t$ in $N$ cycles of duration $\Delta t$, and each $\Delta t$ in 
sub-intervals of length $\lambda_k\Delta t$, that is, $t \equiv N \Delta t$, and $\Delta t \equiv \sum_k \lambda_k\Delta t$. 
At the beginning of the first sub-interval we apply a unitary $V_1$ on $S$, at the second $V_2V_1^\dag,$ at the third 
$V_3V_2^\dag$, and so on. If $N$ is sufficiently large (formally, $N\rightarrow \infty$), the effective Hamiltonian over 
each cycle is simply the time average, namely, 
\[\overline{H}^{(0)} = 
\frac{1}{\Delta t}\sum_k \lambda_k\Delta t ( \gamma V^\dag _k\Pi_S V_k) \otimes X_E=\gamma P\otimes X_E,\]
with the leading-order correction $|| \overline{H}^{(1)}|| = {\mathcal O}[(\Delta t \,|| \gamma\Pi_S\otimes X_E ||)^2].$
Over the time duration $t$, this results in the joint evolution:
\beq
\label{eq:Uent}
\tilde U_{SE}(t) \approx e^{-i \overline{H}^{(0)} t} = \cos(\gamma t P)\otimes I_E +i \sin(\gamma t P)\otimes X_E.
\eeq

\item[(II)] In order to obtain the Kraus operators in the decomposition \eqref{eq:polar}, we next to apply a conditional 
unitary $U_{\rm cond} \equiv U_0\otimes \ket{0}\bra{0}+U_1\otimes\ket{1}\bra{1}$. Once $E$ is traced out, the net 
dynamics on $S$ is a CPTP map with operators 
\( M_0 \approx U_0\cos(t\gamma P),$  $M_1\approx U_1\sin(t\gamma P),\) as desired.
\end{enumerate}

\subsubsection{Beyond rank-two channels.} 
In order to implement CPTP maps of higher rank by using the specified coherent-control resources, we needs 
more copies of the pure auxiliary qubit -- or, as in the original scheme, the ability to dissipatively reset it to the initial pure state, which however is not viable in the present setting.

The basic idea is the following: In order to obtain a map Kraus rank three, associated to Kraus operators 
$M_0,M_1,M_2,$ we first use the procedure described above to implement $M_0$ and an intermediate operator 
$\tilde M_1 \equiv \sqrt{M_1^\dag M_1 +M_2^\dag M_2}.$ Then, conditionally on the state $\ket{1}$ of the first 
ancilla qubit, we performs another unitary design, associated to a CPTP map of Kraus rank 2, but now with 
Kraus operators $M_0' \equiv M_1\tilde M_1^{-1}$, $M_1' \equiv M_2\tilde M_1^{-1}$. This is still a TP map, 
since we have:
\beqan 
M_0'^\dag M_0'+ M_1'^\dag M_1' &=& \tilde M_1^{-1}M_1^\dag M_1\tilde M_1^{-1}+\tilde M_1^{-1}M_2^\dag M_2\tilde M_1^{-1}\\
&=& \tilde M_1^{-1}(M_1^\dag M_1+M_2^\dag M_2)\tilde M_1^{-1} = I.
\eeqan
By tracing out $E$, this leaves a CPTP map on the system, that approximates the target map of rank 3.
Such a ``nested'' construction can be iterated to a general rank $m$ in principle. In this case, the number of auxiliary qubits that are needed is $m-1$. In terms of the dimension of the auxiliary quantum resources that are employed, and depending on the system dimension $d_S$, this procedure is generally inefficient with respect to the minimal Stinespring construction, and more so with respect to the probabilistic design of Sec. \ref{sec:prob}; yet, it has the advantage of providing a systematic 
approach to construct the needed Kraus operators by following the algorithm. 

\subsubsection{Noisy ancilla.} 
It is worth investigating, at least in the simplest, rank-two case, what are the implications of relaxing the 
assumption that the auxiliary environment is purely initialized. Let us assume that the initial state for $E$ is 
mixed, i.e., has the general form
$\rho_E=w_0\ket{0}\bra{0}+w_1\ket{1}\bra{1} + q\ket{0}\bra{1} +q^*\ket{1}\bra{0}.$ By following the same 
evolution of Eq. \eqref{eq:Uent} as in the pure-state case, the joint system-environment state after time $t$ is:
\beqan
\rho_{SE}(t)&=&\tilde U_{SE}(t) \rho_S\otimes\rho_E \tilde U_{SE}^\dag (t)\\
&=&\cos(\gamma t P)\rho_S\cos(\gamma t P)\otimes (w_0\ket{0}\bra{0}+w_1\ket{1}\bra{1})\\&&+ \sin(\gamma t P)\rho_S\sin(\gamma t P)\otimes (w_1\ket{0}\bra{0}+w_0\ket{1}\bra{1})\\
&&+ [\mbox{off-diagonal terms in } \rho_E].
\eeqan
Thus, the reduced states, conditional on $E$ being in $\ket{0}$ or $\ket{1},$ respectively, become :
\[\rho_S|_{\ket{0}}=w_0\cos(\gamma t P)\rho_S\cos(\gamma t P)+w_1\sin(\gamma t P)\rho_S\sin(\gamma t P),\]
\[\rho_S|_{\ket{1}}=w_0\sin(\gamma t P)\rho_S\sin(\gamma t P)+w_1\cos(\gamma t P)\rho_S\cos(\gamma t P).\]
Since these are non-trivial convex combination of single-operator CP maps, in contrast with the pure case 
(which is recovered by posing $w_0=1,w_1=0$), we cannot exploit the polar decomposition in 
\eqref{eq:polar} and obtain the desired reduced dynamics by applying a conditional unitary \footnote{
Interestingly, the above issue does not occur for a rank-2 CPTP map whose Kraus operators are 
{\em  Hermitian and positive-semidefinite} (hence, associated to a unital map), 
as no unitary is needed in the polar decomposition.
However, the method we outlined would still incur in problems for higher-rank Hermitian-Kraus maps, 
since a unitary evolution conditional on the auxiliary state is needed then.}.
If we do apply the same conditional operation of the pure-ancilla case, the same contraction argument of the previous sections 
holds, and the error in the final implementation can be bounded in the $\|\cdot\|_{1\rightarrow 1}$ norm, depending on $w_1$. 
Therefore, with a non-pure ancilla, arbitrary accuracy {\em can no longer} be achieved even in the limit of 
arbitrarily fast control and perfect averaging.

\subsection{Coherent implementation of feedback-decoupling for quantum memory}

In principle, the joint unitary controllability assumption implies that we can obtain any unitary dynamics on the system -- in 
particular, the {trivial one}, corresponding to engineering a quantum memory, or a NOOP gate. In practice, however, it 
may be hard to find explicit controls that enact it while complying with practical constraints, 
and that can ensure robust performance with respect to partial knowledge and uncertainty 
about the coupling with the bath and its internal dynamics.

While open-loop dynamical decoupling techniques offer a method of choice in many quantum information settings of interest 
\cite{viola-dd,viola-memory}, in Ref. \cite{ticozzi-feedbackDD} we presented a way to remove the effect of unwanted environmental 
interactions and effectively decouple $S$ from $B$ by combining coherent-control capabilities with measurement-based single-bit 
feedback. Despite being less flexible in regard to the types of system-bath interactions that are able to be suppressed, 
feedback-enacted decoupling may offer important advantages on time-scale requirements and compensate for 
uncorrelated noise, unlike open-loop schemes. Recently, the method has been successfully demonstrated 
by using a fully coherent implementation to achieve a NOOP gate on a nitrogen-vacancy qubit device in the presence 
of dephasing noise \cite{cappellaro-feedbackDD}.  Here, we recast the original single-bit feedback strategy in the unitary 
design framework, focusing on the {\em exact} correction of an unwanted evolution $U_{SB}$ at a target (finite) time $T$. 
A related strategy for approximate (first-order, short-time) suppression of the corresponding Hamiltonian generator 
$H_{SB}$ is proposed in the original work, and can also be adapted to the present framework along similar lines. The 
protocol may be described as follows: 

\begin{description}
\item[Task] Enact a NOOP gate on the target system $S$ at time $T,$ by removing the net effect of its interaction 
with the uncontrollable bath $B$, so that an arbitrary initial state $\rho_S\equiv |\psi\rangle\langle \psi|$ is preserved.

\item[Environment] The environment comprises both an auxiliary two-level system, with state space $\Hi_{A}=\span\{\ket{0},\ket{1}\}$, 
initially in a pure state $\rho_A=\ket{\phi}\bra{\phi},$ where $\ket{\phi}\equiv (\ket{0}+\ket{1})/\sqrt{2}$, and an uncontrollable bath 
in an arbitrary initial state $\rho_B.$

\item[Free dynamics] $H_0 = H_S \otimes I_B+  I_S\otimes H_B + H_{SB}$, where $H_S$ and 
$H_B$ are arbitrary, and the unwanted coupling Hamiltonian $H_{SB} \equiv S_0\otimes B_0.$
In order for the method to work, we need $S_0$ to have a certain eigenvalue structure, that makes it resemble 
a generalized Pauli matrix -- the precise form is given in the protocol below.

\item[Control resources] We need fast (ideally, instantaneous) conditional unitary transformations of the form:
\[U_0\otimes \ket{0}\bra{0}+U_1\otimes\ket{1}\bra{1} ,\] 
where $U_{0,1}$ are arbitrary unitaries on the system, and an identity action is understood on the bath 
\footnote{In the original proposal \cite{ticozzi-feedbackDD}, some of these 
conditional unitaries are substituted by von Neumann measurement and feedback, implementing unitary evolutions on $S$ alone, 
conditionally on the output of the measurement. }. 
\end{description}

\begin{figure}[t]
\hspace*{20mm}
\includegraphics[width=0.8\columnwidth]{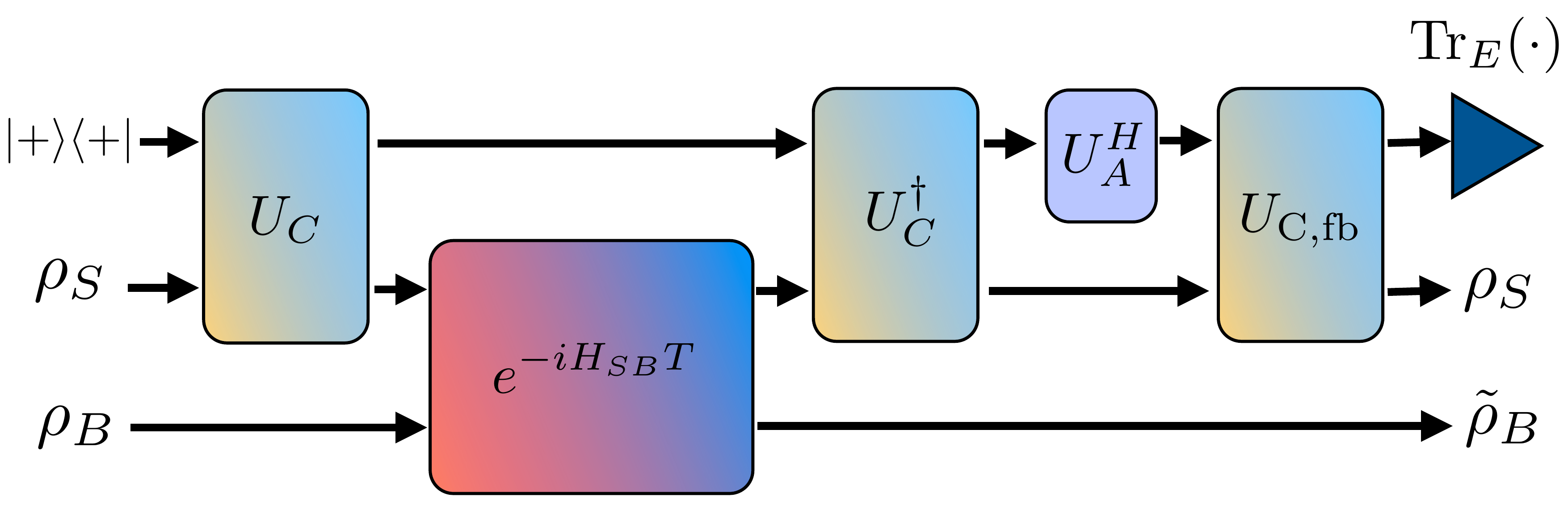}
\vspace*{-3mm}
\caption{(Color online) Coherent feedback averaging protocol. } 
\label{fig:feedbackdd}
\end{figure}
The basic steps of the protocol are (see also Figure \ref{fig:feedbackdd}):

\begin{enumerate}
\item[(I)] Rapidly entangle ${S}$ and ${A}$, by performing a
conditional gate of the form $U_C \equiv \ket{0}\bra{0}\otimes I_S+\ket{1}\bra{1}\otimes U_S$.

\item[(II)] Let the system evolve freely in the presence of ${B}$, up to time
$T$, under the joint unitary propagator $U_{SB}(T) =\exp(-i H_{0}T). $

\item[(III)] Apply a second fast conditional unitary, the inverse of the one in step (I), 
that is: $U_C^\dag=\ket{0}\bra{0}\otimes I_S+\ket{1}\bra{1}\otimes U_S^\dag.$

\item[(IV)] Apply a Hadamard transformation $U^H_{A}$ \cite{nielsen-chuang}
to the ancilla qubit ${A}$.
\end{enumerate}

\noindent 
The resulting joint state on the whole system at this stage may be written as:
\beqan
\rho_{SE}(T)&=\ket{0}_A\bra{0}\otimes\left(A_{SB}^{(+)}(T)
\rho_{SB}(0) A^{(+)\dagger}_{SB}(T) \right)\\
&
+\ket{1}_A\bra{1}\otimes\left(A_{SB}^{(-)}(T)
\rho_{SB}(0) A^{(-)\dagger}_{SB}(T) \right)\,,\eeqan
\noindent 
where $\rho_{SB}(0)= |\psi\rangle\langle \psi| \otimes \rho_B$ and 
\beq 
A^{(\pm)}_{SB}(T) =\frac{1}{2}\left( U_{SB}\pm
(U^\dag_S \otimes I_E) U_{SB}(U_S \otimes I_E)\right)\,,
\end{equation}
The form of $A_{SB}^{(+)}(T)$ shows that the evolution conditional to $\ket{0}$ on $A$ already enacts, at time $T$, 
an {\em exact average} of the unitary propagators 
 $U_{SB}$ and $(U^\dag_S \otimes I_E) U_{SB} (U_S\otimes I_E)$. To our aim, such an average should return an 
 operator of the form $I_S\otimes B_1,$ with $B_1$ an arbitrary operator on the bath, thus yielding a trivial evolution on 
 the target system, as desired. In Ref. \cite{ticozzi-feedbackDD}, we show that $U_S$ can achieve this task if and only 
 if $U_{SB}$ satisfies the following property: when it is written as a block-matrix, accordingly to the tensor structure of 
 $\Hi_B\otimes\Hi_S$ (notice the swap of the two factors, for convenience in analyzing its block structure), its 
 $d_S\times d_S$ blocks $U_{SB}(i,j)$ are of the form:
\beq
\label{eq:blockform} 
U_{SB}(i,j)=a_{i,j}I_S + b_{i,j}X,
\eeq
where $X$ is a normal matrix whose eigenvalue $x_k$, in decreasing order, must further satisfy a {\em mixing condition}, namely, 
for all $k,$ $x_k=-x_{d_S-k}.$ 

The last required step in the protocol aims to transform $A_{SB}^{(-)}(T)$ into an operator of the form $I_S\otimes B_2$ as well, 
ensuring trivial dynamics on the system of interest irrespective of the auxiliary system.
If $U_{SB}$ has the block-form \eqref{eq:blockform}, then it is easy to see that the corresponding blocks for $A^{(-)}_{SB}$  are of the form
\( A^{(-)}_{SB}(i,j)= c_{i,j}X.\)
Different blocks can be made proportional to one another, via unitary conditional operations, if and only if $X$ is itself proportional to a unitary, 
say, $U^\dag_{\rm fb}$. Then the following conditional operation will correct the ``wrong'' averaging implemented by 
$A^{(-)}_{SB}(T)$:

\begin{enumerate}
\item[(V)] Apply a third conditional unitary $U_{\rm C-fb} \equiv \ket{0}\bra{0}\otimes I_S+\ket{1}\bra{1}\otimes U_{\rm fb}.$
\end{enumerate}

By explicitly writing $U_{SB}(T) =\exp[-i (H_S\otimes I_B+I_S\otimes H_B + S_0 \otimes B_0)T],$ it is easy to see that blocks 
$U_{SB}(i,j)$ in Eq. \eqref{eq:blockform} will be linear combinations of $I_S$ and powers of $S_0.$ If $S_0^2=I,$ and $S_0$ satisfies 
the mixing condition stated above, then feedback decoupling is 
possible. More explicitly, $S_0$ can be corrected if, up to a change of basis in $S$ and a reshuffling of its eigenvalues, 
we have $S_0=\sigma_z\otimes I_{d_S/2}.$  Physically, this means that the noise induced by the unwanted system-bath coupling 
is purely dephasing, and further subject to the above symmetry constraint.
When the target system is a qubit, $d_S=2$, as for instance in the experimental implementation of \cite{cappellaro-feedbackDD}, 
this constraint is automatically satisfied in the pure dephasing regime.

\subsection{Coherent implementation of splitting-subspace approach for quantum stabilization}

The preparation of a target quantum state in a system of interest, in a way that is independent with respect to its initial state, 
is a key task in quantum control, motivated by applications ranging from quantum information processing 
 to quantum purification and cooling, see e.g. 
\cite{kraus-dissipative, ticozzi-QDS, wolf-sequential, Verstraete2009,ticozzi-cooling} and references therein.
If the target state is also required to be {\em invariant} (a fixed point) for the underlying dynamics, then from a system-theory 
standpoint the task becomes one of {\em stabilization}. This offers the important advantage of not only generating 
(either asymptotically or in finite time) the quantum state of interest; in addition, it can also maintain it in a way that is 
insensitive to certain types of errors and uncertainties -- as we discussed in detail in Refs. \cite{ticozzi-alternating,johnson-FTS}. 
It is clear that, in order for $S$ to converge towards its steady state or, more generally, a steady-state subspace, by ``forgetting'' 
its initial condition, the evolution on $S$ must be irreversible. The required stabilizing continuous-time or 
discrete-time dynamics may be synthesized either in a purely open-loop fashion or, most commonly, by relying on 
measurement-based or coherent quantum feedback with a suitable auxiliary system (see e.g. \cite{altafini-tutorial} 
for an overview). 

From the perspective of unitary design, we showed in Ref. \cite{ticozzi-cooling} how the resources needed to exactly 
stabilize a target pure state must include a purely-initialized virtual subsystem of appropriate dimension. Here, 
building on a constructive approach introduced in Ref. \cite{baggio-CDC}, we demonstrate how to use a sequence of 
unitary operations to stabilize an arbitrary pure state or a subspace on $S$ in finite time. With respect to the original 
proposal, which involved repeated uses of a single ancillary qubit, we allow for a larger auxiliary system, 
in the same spirit of Sec. \ref{subsec:lv}.
This, in turn, enables us to avoid the need for a dissipative resetting operation and obtain the desired output 
effectively in a single step. The protocol may be described as follows:

\begin{description}

\item[Task] Given a target subspace $\Hi_T \subseteq\Hi_S$ 
of dimension $d_T,$ enact a CPTP map $\Ti = \Ec_{T,0}$ on $S$ so that $\Ec_{T,0}(\rho_S)\in{\mathfrak D}(\Hi_T)$ for any 
$\rho_S\in{\mathfrak D}(\Hi_S)$. This is equivalent to requiring $\Ec_{T,0}(\rho_S)\leq \Pi_T$ for all initial states, 
where $\Pi_T$ is the orthogonal projection onto $\Hi_T.$

\item[Environment] An auxiliary system $\Hi_E \equiv \span\{\ket{1},\ldots,\ket{K}\}$, where $K$ is the first integer 
greater than $d_S/d_T$, initially in a pure state, say, $\rho_E\equiv \ket{1}\bra{1}.$ 
No interaction with an uncontrollable bath is considered.

\item[Control resources] With respect to $\Hi_A \otimes \Hi_S$, we need fast (ideally, instantaneous) conditional 
unitaries of the form:
\beq 
U_{C,S} \equiv \sum_k U_k\otimes \Pi_k,
\label{Ucs}
\eeq
where the orthogonal projectors $\Pi_k$ are a resolution of the identity on $\Hi_S$, to be specified later, 
and $U_k$ are unitary operations on $\Hi_A$, such that $U_k\ket{1}=\ket{k}$. Additionally, we need 
conditional unitaries of form 
\beq
U_{C,A} \equiv \sum_k \ket{k}\bra{k}\otimes V_k,
\label{Uca}
\eeq
where $V_k$ are unitaries on $S.$ 
\end{description}

\begin{figure}[t]
\hspace*{27mm}
\includegraphics[width=0.76\columnwidth]{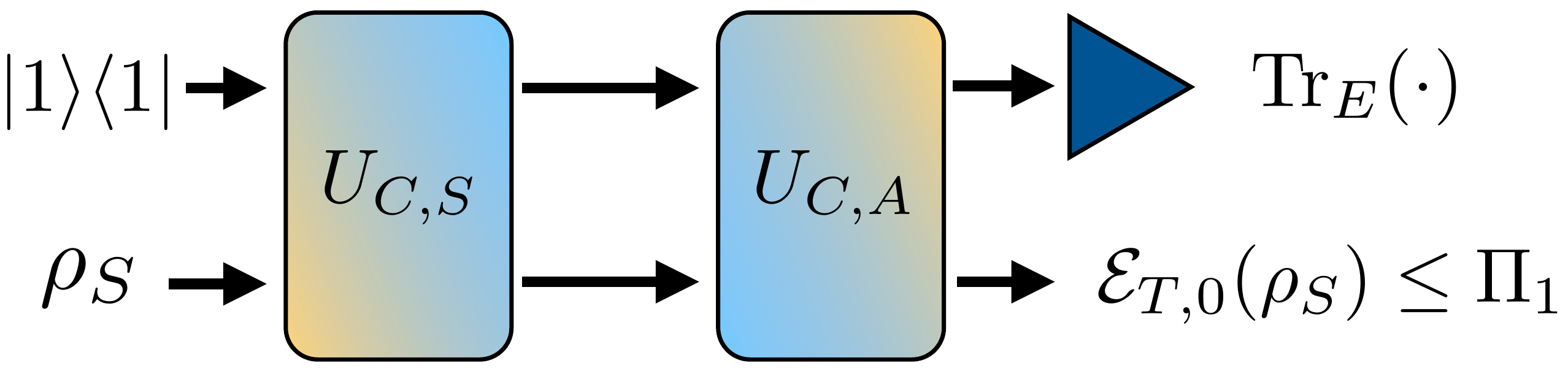}
\vspace*{-3mm}
\caption{(Color online) Coherent implementation of subspace stabilization protocol. } 
\label{fig:splitting}
\end{figure}

The starting point of the procedure is to decompose $\Hi_S$ as the direct sum 
\[\Hi_S \equiv \bigoplus_{k=1}^K\Hi_{k},\]
where the subspaces $\Hi_k$, $k=1,\ldots,k-1$, are isomorphic orthogonal copies of $\Hi_T$, with 
$\Hi_1\equiv\Hi_T,$ and $\Hi_{K}=\Hi_S\ominus\left( \bigoplus_{k=1}^{K-2}\Hi_k \right).$
The idea is to use $U_{C,S}$ to ``encode'' in the state of the auxiliary system the information regarding {\em which} 
of the subspaces $\Hi_k$ is populated, and then to use $U_{C,A}$ in order to obtain a final state that 
populates {\em only} the target one. To do so, we choose $\{\Pi_k\}$ in Eq. \eqref{Ucs} as the orthogonal 
projections onto the corresponding $\Hi_k.$  In addition, we choose $V_k$ in Eq. \eqref{Uca} such that 
$V_k\Pi_k V_k^\dag \leq \Pi_1.$ Notice that this is possible if and only if $\rank(\Pi_k) \geq \rank(\Pi_1)$ 
for all $k$, as is the case with the $\Pi_k$ chosen as above.   
Formally, this simple protocol then consists of the following steps:
\begin{enumerate}
\item[(I)] Apply $U_{C,S}$;

\item[(II)] Apply $U_{C,A}$.
\end{enumerate}
\noindent 
It is a matter of direct calculation to verify that the final evolution is:
\beq
\label{eq:cohfb}
\Ec_{T,0}(\rho_{S})=\sum_{k=1}^K V_k\Pi_k\rho_S\Pi_k V_k^\dag, 
\eeq
and $\Ec_{T,0}(\rho_{S})\leq\Pi_1=\Pi_T$ by construction. 
The protocol is illustrated schematically in Fig. \ref{fig:splitting}. A few remarks are in order:

\begin{itemize}

\item Eq. \eqref{eq:cohfb} shows that such a splitting-subspace approach provides a general way to ``embed'' a 
measurement-based, discrete-time feedback protocol, such as those described in Ref. \cite{bolognani-arxiv}, 
within a coherent-feedback picture. In fact, the same CPTP evolution of Eq. \eqref{eq:cohfb} may be obtained as the 
average outcome of the following procedure: first, perform a projective measurement of an observable with spectral decomposition 
$O=\sum^{K}_{k=1} \lambda_k\Pi_k$; next, apply a unitary evolution $V_k$ conditional on the $k$th outcome of the measurement.
 
\item When the target is a pure state, $\Hi_T \equiv \span \{\ket{\psi}\bra{\psi}\}$, in the above protocol we need to choose 
$K=d_S$, which corresponds to the Kraus rank of the extreme CPTP map that realizes the target all-to-one evolution, 
namely, ${\cal T}(\rho_S)=\ket{\psi}\bra{\psi}\Tr(\rho_S).$ The amount of resources requested by this coherent 
implementation is thus {\em optimal}.  Optimality remains true 
when the target is a subspace, and $d_S$ is a multiple of $d_T$. This may be seen by recasting the problem 
as the stabilization of the pure state $\ket{1}\bra{1}$ of a 
virtual subsystem $\Hi_V$, defined via the decomposition $\Hi_S \equiv \Hi_V\otimes \Hi_T
\equiv \span\{\ket{k},k=1,\ldots,K\}\otimes\Hi_T.$ Then the same necessary conditions about exact pure-state preparation 
ensure that a pure ancilla of dimension $K=d_S/d_T$ is precisely the minimal auxiliary resource that allows for 
engineering the target. We expect a similar result to hold more generally, when $d_T$ does not divide $d_S$.

\item While in the coherent-feedback loop described above we have only considered {\em exact} stabilization (hence, 
an auxiliary system in a pure state), it is straightforward to extend the method to the case where {$\rho_E$ is mixed}, 
as long as $E$ contains a virtual subsystem of sufficient dimension initialized in a pure state. 
Since the resulting evolution a trace-norm contraction, if the initialization is only $\varepsilon$-approximate  
the final state will be at most $\varepsilon$-distant from the intended target.

\item The possibility of stabilizing subspaces in one step allows us to envision a coherent-feedback implementation of the 
quasi-local dissipative circuits introduced in Refs. \cite{johnson-FTS} to stabilize entangled pure states in finite time, 
{\em robustly} with respect to the order of the applied maps. Such circuits can be constructed when the target state 
is able to be represented as a product state with respect to a suitable, locality-constrained decomposition in virtual 
subsystems of the multipartite system. In this setting, in order to implement the quasi-local stabilizing maps, we 
would need to couple the target system to an auxiliary system  consisting of isomorphic copies of each virtual subsystem -- 
each prepared in a known pure state, in a way that respects the specified locality constraints.

\end{itemize}
 
\section{Outlook}

We have presented a mathematical characterization of the resources needed to engineer CPTP open quantum dynamics based on a
coherent, unitary design approach -- as informed by a 
virtual-subsystem perspective.  While our emphasis has been on establishing general (non-constructive) channel 
controllability results, we have shown how our framework is highly flexible and easily applicable to the description of 
existing constructive schemes for universal open-system simulation or quantum channel construction. In addition to 
the illustrative applications we have discussed, other quantum protocols of interest, which may be well-worth analyzing in our 
unitary-design framework, include {\em collisional and repeated interaction} models for non-Markovian dynamics 
\cite{ziman-simulation,pellegrini-repeated}.
From a control-theoretic standpoint, developing explicit optimal-control algorithms tailored to the equivalent state-transfer 
problem we have associated to quantum-channel synthesis is, as we mentioned, an interesting and natural direction for 
investigation.  Likewise, while the present analysis has relied crucially on the assumption of complete unitary controllability of 
system and environment together, it would be desirable to characterize more general scenarios where only partial controllability is 
assumed -- possibly exploring relationships between exact and approximate controllability notions, in analogy to closed 
quantum systems \cite{Boscain}.

\section*{Acknowledgements}

It is a pleasure to thank David Reeb for insightful discussions that partially motivated the present work.
Work at Dartmouth was supported by the National Science Foundation through grant no. PHY--1620541 and 
the Constance and Walter Burke Special Projects Fund in Quantum Information Science.

\section*{References}


\providecommand{\newblock}{}

\end{document}